\documentclass[conference]{IEEEtran}
\IEEEoverridecommandlockouts
\usepackage{cite}
\usepackage{amsmath,amssymb,amsfonts}
\usepackage{algorithmic}
\usepackage{graphicx}
\usepackage{textcomp}
\usepackage{tabularx}
\usepackage{xcolor}
\usepackage{url}
\usepackage{subfig}

\usepackage{float}
\def\BibTeX{{\rm B\kern-.05em{\sc i\kern-.025em b}\kern-.08em
		T\kern-.1667em\lower.7ex\hbox{E}\kern-.125emX}}
\begin{document}
	
	\title{A PBN-RL-XAI Framework for Discovering a ``Hit-and-Run'' Therapeutic Strategy in Melanoma\\

	}
	
\author{\IEEEauthorblockN{1\textsuperscript{st} Zhonglin Liu}
	\IEEEauthorblockA{\textit{Department of Mathematics, } \\
		\textit{The University of Hong Kong}\\
		Hong Kong, Hong Kong \\
		u3597461@connect.hku.hk}}
	
	\maketitle

	\begin{abstract}
		Innate resistance to anti-PD-1 immunotherapy remains a major clinical challenge in metastatic melanoma, with the underlying molecular networks being poorly understood. To address this, we constructed a dynamic Probabilistic Boolean Network model using transcriptomic data from patient tumor biopsies to elucidate the regulatory logic governing therapy response. We then employed a reinforcement learning agent to systematically discover optimal, multi-step therapeutic interventions and used explainable artificial intelligence to mechanistically interpret the agent's control policy. The analysis revealed that a precisely timed, 4-step temporary inhibition of the lysyl oxidase like 2 protein (LOXL2) was the most effective strategy. Our explainable analysis showed that this ``hit-and-run" intervention is sufficient to erase the molecular signature driving resistance, allowing the network to self-correct without requiring sustained intervention. This study presents a novel, time-dependent therapeutic hypothesis for overcoming immunotherapy resistance and provides a powerful computational framework for identifying non-obvious intervention protocols in complex biological systems.
	\end{abstract}
	

\section{Introduction}

Immune checkpoint blockade targeting the programmed death-1 (PD-1) pathway has revolutionized treatment for metastatic melanoma, yet innate resistance affects 60-70\% of patients \cite{hugo2016genomic}. Unlike acquired resistance that develops during treatment, innate resistance manifests as immediate therapeutic failure, suggesting pre-existing molecular networks that render tumors refractory to immune activation from the outset. The mechanistic basis for this variation in response patterns remains poorly explained, representing a critical unmet clinical need.

Recent transcriptomic analyses have identified gene expression signatures associated with resistance, most notably the Innate anti-PD-1 Resistance (IPRES) signature, which encompasses genes involved in mesenchymal transition, extracellular matrix remodeling, angiogenesis, and wound healing \cite{hugo2016genomic}. However, the regulatory logic governing how these genes interact dynamically to establish and maintain resistant phenotypes remains largely opaque. Traditional reductionist approaches focusing on individual biomarkers have failed to capture the systems-level properties that govern therapeutic response, largely due to the inherent complexity of the underlying gene regulatory networks (GRNs) \cite{kreeger2010cancer}.

Current therapeutic discovery approaches face fundamental limitations when applied to complex resistance networks. The vast combinatorial space of possible multi-gene interventions makes exhaustive experimental screening prohibitive. Moreover, optimal intervention strategies may involve precisely timed, transient perturbations rather than sustained pathway inhibition \cite{cornelius2013realistic}, and the stochastic nature of biological networks necessitates probabilistic control strategies.

While existing computational approaches for therapeutic discovery have explored various methodologies—including Boolean network control \cite{murrugarra2015molecular}, dynamical systems optimization \cite{alaimo2018network}, and machine learning-based drug repurposing \cite{pushpakom2019drug}—these methods typically focus on static network analysis or sustained interventions. Our approach uniquely integrates Probabilistic Boolean Networks (PBNs) with reinforcement learning (RL) to discover time-dependent, transient intervention strategies. PBNs combine computational tractability with biological realism through stochastic dynamics \cite{shmulevich2002probabilistic}, while RL enables discovery of optimal control policies without requiring explicit system knowledge \cite{sutton1998reinforcement}. This integration allows systematic exploration of dynamic intervention strategies that existing static approaches cannot capture.

Here, we present this novel computational framework to systematically discover optimal therapeutic interventions for overcoming innate anti-PD-1 resistance in melanoma. We constructed dynamic PBN models from patient transcriptomic data to capture the regulatory logic governing therapy response, then employed RL agents to identify multi-step intervention strategies. Using explainable artificial intelligence (XAI), we mechanistically interpreted the learned control policies to understand their biological basis.

Our analysis reveals that innate resistance corresponds to a fundamental system-level transition from a plastic, multi-stable network state to a rigid, canalized landscape dominated by a JUN/LOXL2 regulatory axis. Most significantly, our RL approach discovered a non-obvious ``hit-and-run" therapeutic strategy: a precisely timed, 4-step temporary inhibition of LOXL2 achieving 93.45\% success rate in silico. This computational hypothesis requires rigorous experimental validation before clinical consideration, yet demonstrates the framework's potential for discovering non-intuitive therapeutic strategies in complex biological systems.

	\section{Methods}
	
\subsection{Data Acquisition and Binarization}

We utilized the landmark RNA-sequencing dataset from Hugo et al. \cite{hugo2016genomic}, publicly available from the Gene Expression Omnibus\cite{clough2016gene} (accession: GSE78220). This dataset comprises transcriptomic profiles from 28 pre-treatment melanoma biopsies obtained from patients undergoing anti-PD-1 checkpoint inhibitor therapy (pembrolizumab or nivolumab) for metastatic melanoma. The cohort included patients aged 19-84 years (median 60 years) with 75\% male representation and predominantly advanced disease (89\% M1b/M1c stage). Approximately 50\% had prior MAPK inhibitor treatment.

Pre-treatment tumor biopsies were stratified into ``Responders" (n=15), defined as patients achieving Complete or Partial Response, and ``Non-Responders" (n=13), defined as patients with Progressive Disease according to immune-related response criteria. One on-treatment sample was excluded to ensure analysis focused exclusively on the innate resistance state. Gene identifiers were mapped to official HUGO Gene Symbols for biological interpretation.

To prepare data for Boolean modeling, continuous gene expression values were converted into discrete binary states of 0 (OFF/low activity) or 1 (ON/high activity). Raw count data were filtered to include only core genes, then Variance Stabilizing Transformation (VST) \cite{durbin2002variance} was applied using DESeq2 \cite{love2014moderated} to stabilize variance across expression ranges.

For each gene within each cohort's VST-normalized matrix, we employed a Gaussian Mixture Model (GMM) \cite{reynolds2015gaussian} to determine data-driven binarization thresholds. The probability density of expression value $x$ was modeled as:
\begin{equation}
	p(x) = \pi_1 \mathcal{N}(x | \mu_{\text{low}}, \sigma^2) + \pi_2 \mathcal{N}(x | \mu_{\text{high}}, \sigma^2)
	\label{eq:gmm}
\end{equation}
where $\pi_1$ and $\pi_2$ are mixing coefficients summing to 1, and $(\mu_{\text{low}}, \mu_{\text{high}}, \sigma^2)$ represent component means and variance. Parameters were fitted using mclust R package, with binarization threshold $T = (\mu_{\text{low}} + \mu_{\text{high}}) / 2$. This yielded two binarized matrices for subsequent network inference.

\subsection{Introduction of PBN}

To capture the complex and stochastic nature of gene regulatory networks (GRNs), we employ the Probabilistic Boolean Network (PBN) framework \cite{shmulevich2002probabilistic}. A PBN models $n$ genes as binary variables with state vector $\mathbf{x}(t) = (x_1(t), \ldots, x_n(t))$ where $x_i(t) \in \{0, 1\}$ represents gene $g_i$'s state at time $t$.

For each gene $g_i$, there exists a set of $k_i$ Boolean predictor functions $F_i = \{f_1^{(i)}, \ldots, f_{k_i}^{(i)}\}$. At each time step, function $f_j^{(i)}$ is selected with probability $c_j^{(i)}$ where $\sum_{j=1}^{k_i} c_j^{(i)} = 1$, determining the next state: $x_i(t+1) = f_j^{(i)}(\mathbf{x}(t))$.

The long-term behavior is characterized by attractors—minimal sets of states $A$ where $P(\mathbf{x}(t+1) \in A | \mathbf{x}(t) \in A) = 1$ and $P(\mathbf{x}(t+1) \notin A | \mathbf{x}(t) \in A) = 0$. In systems biology, attractors represent stable cellular phenotypes such as therapy-resistant states \cite{shmulevich2002boolean}. 

To quantify regulatory relationships, we compute the influence of gene $g_j$ on $g_i$ as the expected sensitivity across all predictor functions:
\begin{equation}
	I(g_j \to g_i) = \sum_{k=1}^{k_i} c_k^{(i)} \cdot S(f_k^{(i)}, g_j)
	\label{eq:influence}
\end{equation}
where $S(f_k^{(i)}, g_j)$ is the sensitivity of function $f_k^{(i)}$ to gene $g_j$. The resulting influence matrix provides a quantitative map of the network's regulatory structure.

\subsection{PKN-constrained PBN construction}

PBN inference is computationally intractable for large networks due to double-exponential growth in search space. We developed a multi-stage procedure to construct a tractable, biologically-relevant 12-gene core network from melanoma anti-PD-1 resistance data.

We began with the KEGG pathway hsa05235 ``PD-L1 expression and PD-1 checkpoint pathway" \cite{kanehisa2000kegg}, directly relevant to anti-PD-1 therapy mechanisms. Our hybrid selection strategy mandatorily included five IPRES signature genes (AXL, ROR2, WNT5A, LOXL2, TAGLN) \cite{hugo2016genomic} known to drive therapy resistance through mesenchymal transition and angiogenesis pathways. The remaining seven genes were selected using a composite score combining differential expression significance ($S_{DE}$) calculated as $-\log_{10}(p_{adj})$ from Benjamini-Hochberg adjusted p-values \cite{haynes2013benjamini} from differential expression analysis \cite{anders2010differential}, and network centrality ($S_{Net}$) from dynGENIE3 inference \cite{huynh2018dyngenie3}:
\begin{equation}
	S_{final} = 0.5 \cdot S_{DE,norm} + 0.5 \cdot S_{Net,norm}
\end{equation}

The 12-gene limitation balances computational tractability ($2^{12} = 4096$ states) with biological comprehensiveness. The selected genes span key resistance mechanisms: immune checkpoint regulation (PDCD1, CD274), mesenchymal transition (AXL, ROR2, WNT5A), extracellular matrix remodeling (LOXL2, TAGLN), and metabolic reprogramming. Prior studies demonstrate that core regulatory modules of 10-15 genes can capture essential cellular dynamics \cite{shmulevich2002probabilistic}.

For PBN inference, we constrained potential regulators for each target gene using dynGENIE3 rankings, testing regulator sets of size $k = 1$ to $k_{max} = 4$. The optimal $k^*$ was selected by maximizing Mutual Information (MI) \cite{kraskov2004estimating}. For each gene's probabilistic update rule, we selected the top $N=4$ Boolean functions, with probabilities assigned proportionally:
\begin{equation}
	P(f_i) = \frac{MI(f_i)}{\sum_{j=1}^N MI(f_j)}
	\label{eq:prob_func}
\end{equation}
We also evaluated Coefficient of Determination (CoD) \cite{ma2009probabilistic} as an alternative metric, but MI proved superior due to CoD's susceptibility to negative values in our dataset.. This procedure generated two context-specific PBN models for Responder and Non-Responder cohorts, capturing distinct regulatory logic underlying therapy response patterns.

\subsection{Reinforcement Learning for Optimal Dynamic Control}

To discover optimal intervention strategies in the resistant-state PBN, we formulated the control problem as a Markov Decision Process (MDP) \cite{puterman1990markov} using the gym-PBN framework \cite{mizera2025pbn}. The state space $\mathcal{S}$ comprises $2^{12} = 4096$ binary network states, with action space $\mathcal{A}$ including ``Do Nothing" and single-gene state flips. Transition probabilities $P(\mathbf{x}_{t+1} | \mathbf{x}_t, a_t)$ are defined by PBN update rules under interventions.

The reward function incentivizes therapeutic transitions: $+100$ for reaching sensitive states, $-5$ for resistant attractors, and $0$ action cost to encourage active intervention within the 15-step clinical horizon. This design reflects the clinical priority of escaping resistance over minimizing intervention frequency.

We employed Proximal Policy Optimization (PPO) \cite{schulman2017proximal} for its proven stability in complex biological control tasks. PPO optimizes the clipped surrogate objective:
\begin{equation}
	L^{CLIP}(\theta) = \mathbb{E}_t \left[ \min \left( r_t(\theta) A_t, \text{clip}(r_t(\theta), 1-\epsilon, 1+\epsilon) A_t \right) \right]
\end{equation}
where $r_t(\theta) = \frac{\pi_{\theta}(a_t | \mathbf{x}_t)}{\pi_{\theta_{old}}(a_t | \mathbf{x}_t)}$ represents the policy ratio and $A_t$ the advantage estimate.

We employed standard PPO hyperparameters from the Stable-Baselines3 ~\cite{raffin2021stable} implementation: discount factor $\gamma = 0.99$ for long-term reward optimization, GAE parameter $\lambda = 0.95$ to balance bias-variance in advantage estimation, and clipping parameter $\epsilon = 0.2$ to prevent destabilizing policy updates while maintaining exploration. These hyperparameters were determined through preliminary experiments to ensure stable performance on our PBN model..

\subsection{Sequential Intervention and Explainable AI Analysis}

\subsubsection{Optimal ``Hit-and-Run'' Strategy Identification}
We designed a sequential experiment to test clinically translatable transient interventions, where brief perturbations destabilize resistant states without sustained therapy. The protocol consisted of two phases: a \textit{priming phase} with temporary target gene inhibition (LOXL2, MAPK3, or JUN clamped to 0 for 1-5 steps), followed by a \textit{control phase} where the pre-trained PPO agent guided the primed network toward sensitive states within the remaining 15-step episode. Success rates identified optimal transient intervention strategies.

\subsubsection{Mechanistic Policy Explanation via SHAP}
We employed SHAP (SHapley Additive exPlanations) \cite{lundberg2017unified} to analyze agent decision-making by quantifying each gene's contribution to intervention choices through game-theoretic marginal contribution analysis. SHAP analysis identified context-dependent vulnerability signatures triggering specific interventions and assessed how transient priming altered these signatures. Trajectory visualizations tracked feature importance evolution across control episodes for dynamic interpretation of agent reasoning.

	\section{Results and Discussions}

\subsection{Network Topology Rewiring Drives Attractor Landscape Rigidification}

To investigate how network rewiring affects system dynamics, we performed computational attractor analysis on both PBN models using the \texttt{optPBN} MATLAB toolbox\cite{trairatphisan2014optpbn}. Each PBN was sampled as 1,000,000 individual Boolean networks to map the probability distribution of steady-state and cyclical attractors.

Network topology comparison reveals fundamental structural differences between responder and non-responder networks (detailed network visualizations in Appendix Fig.~\ref{fig:network_topology}). The non-responder network exhibits higher connectivity (45 vs 41 edges, density 0.341 vs 0.311) with TAGLN functioning as a resistance hub receiving inputs from RELA and JUN while regulating ROR2. Conversely, the responder network shows simplified TAGLN regulation and distinct AXL pathway rewiring toward LOXL2-mediated ECM remodeling. 

Quantitative analysis of node-level topology differences (Appendix Fig.~\ref{fig:topology_heatmap}) identifies nodes with substantial changes: absolute in-degree or out-degree differences $\geq$1, or average mutual information differences $\geq$0.2. TAGLN emerges as the most dramatically altered node, with 3-fold reduced in-degree connectivity in responders (-3 inputs, -1 output), confirming its critical role as a resistance coordination hub that becomes disrupted in treatment-responsive networks.

These topological differences drive a dramatic reorganization of the dynamic landscape. The responder network exhibits high plasticity with 22 distinct attractors, where the most probable accounts for only 12.55\% of the total probability basin. In contrast, the non-responder network rigidifies into 16 attractors dominated by a single state capturing 50.02\% of probability mass, indicating robust maintenance of resistance through the TAGLN-coordinated regulatory circuit.

\begin{table}[htbp]
	\caption{Key Cellular Phenotypes Identified Through Attractor Analysis}
	\begin{center}
		\begin{tabular}{|c|c|}
			\hline
			\textbf{Phenotype} & \textbf{(AXL, WNT5A, ROR2)} \\
			\hline
			Responder (Sensitive) & (0, 0, 1) \\
			\hline
			Resistant (WNT-Driven) & (0, 1, 1) \\
			\hline
			Resistant (AXL-Driven) & (1, 0, 1) \\
			\hline
			Dominant Resistant State & (0, 0, 0) \\
			\hline
		\end{tabular}
		\label{tab:phenotype_summary}
	\end{center}
\end{table}

This analysis demonstrates that innate resistance represents a system-level transition from plastic, multi-stable dynamics to a rigid, canalized regime where specific regulatory circuits (particularly the TAGLN-ROR2 axis) become hard-wired to maintain resistance. The quantitative topology analysis provides mechanistic insight into how network rewiring patterns determine therapeutic response outcomes.

	\subsection{Network Rewiring Establishes a JUN/LOXL2 Regulatory Axis}
	\label{sec:rewiring}
	Having established that the resistant state is defined by a rigid and dominant attractor, we next sought to identify the specific regulatory changes responsible for this landscape collapse. We performed a comparative analysis of the PBN models, combining a quantitative assessment of gene influence with a detailed inspection of the inferred Boolean logic.
	
\begin{figure}[t]
		\centering
		\includegraphics[width=\columnwidth]{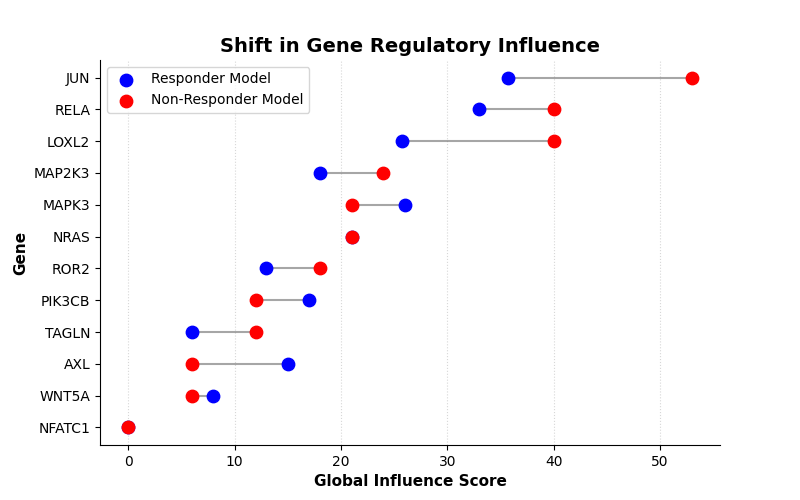}
		\caption{Shift in total gene influence from the responder (blue) to the non-responder (red) model. The plot highlights the change in influence score for each gene, sorted by their final influence in the non-responder state.}
		\label{fig:influence_plot}
	\end{figure}

	To identify the key regulatory hubs, we calculated a \textit{total influence score} for each gene by summing all its incoming and outgoing edge weights from the network's influence matrix. This score quantifies a gene's overall impact as both a regulator and a target within the system. The analysis revealed a significant shift in the network's command structure, establishing the transcription factor JUN as the unequivocal master regulator of the resistant state, with LOXL2 emerging as its primary executive arm. As shown in Fig.~\ref{fig:influence_plot}, JUN's total influence increases dramatically in the non-responder network, and LOXL2 exhibits the second-largest gain. This ascent pinpoints the JUN/LOXL2 axis as the central driver of the switch to the therapy-resistant phenotype.
	
	This dramatic rise in influence is the direct result of a fundamental rewiring of both genes' regulatory logic. In the sensitive responder network, JUN's activation is highly conditional, relying on a specific context of present and absent upstream signals. In stark contrast, the logic in the resistant network becomes decisively canalized, collapsing to a simple, robust AND gate that makes JUN's activation an ``all-or-nothing'' event: \texttt{JUN = MAP2K3 AND NRAS AND RELA AND LOXL2}
	This rewiring creates what is effectively a \textit{hard-wired activation switch} for the master regulator.
	Concurrently, the regulatory logic for LOXL2, the axis's executive arm, undergoes a similar canalization. In the responder network, LOXL2's state is determined by a complex, probabilistic mixture of functions. In the resistant network, this complexity collapses into a single, deterministic rule that enslaves LOXL2 to its new master regulator: \texttt{LOXL2 = JUN AND MAP2K3}
	
	This hierarchical simplification, where JUN is rewired into a master switch and LOXL2 is rewired into its direct effector, hard-wires the JUN/LOXL2 axis and locks the system into a robustly resistant phenotype.
	
\subsection{Temporal Inhibition Reveals a ``Hit-and-Run'' Therapeutic Strategy}
To identify a clinically relevant transient intervention, we simulated temporary ``priming'' inhibition of key targets (JUN, LOXL2, MAPK3) for varying durations, followed by RL-guided control to the therapy-sensitive state.

\begin{figure}[t]
	\centering
	\includegraphics[width=\columnwidth]{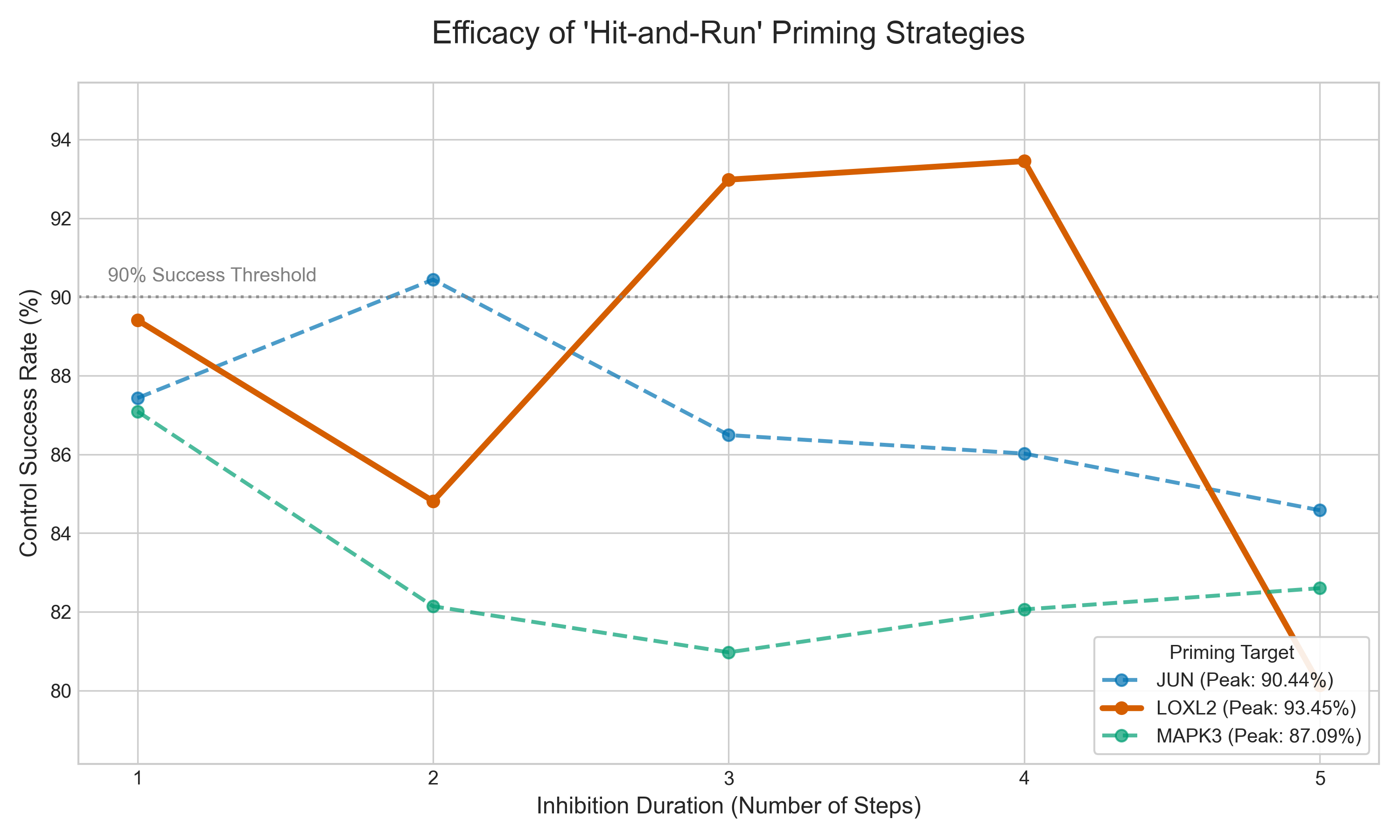}
	\caption{Efficacy of ``hit-and-run'' priming strategies. A 4-step inhibition of LOXL2 emerges as the optimal transient strategy, significantly outperforming other interventions.}
	\label{fig:temporal_plot}
\end{figure}

The results (Fig.~\ref{fig:temporal_plot}) identified an optimal strategy: 4-step LOXL2 inhibition achieving 93.45\% success rate. The non-monotonic relationship between inhibition duration and success for both LOXL2 and JUN demonstrates a ``hit-and-run'' mechanism where precisely timed intervention outperforms sustained targeting. Critically, the agent's overwhelming post-priming policy was ``Do Nothing,'' validating that brief intervention destabilizes resistance, allowing natural dynamics to complete transition.

\subsection{Mechanistic Policy Explanation via SHAP}
We employed SHAP analysis\cite{lundberg2017unified} to understand the agent's control reasoning, providing mechanistic explanations for its strategy.

\begin{figure}[htbp]
	\centering
	\includegraphics[width=\columnwidth]{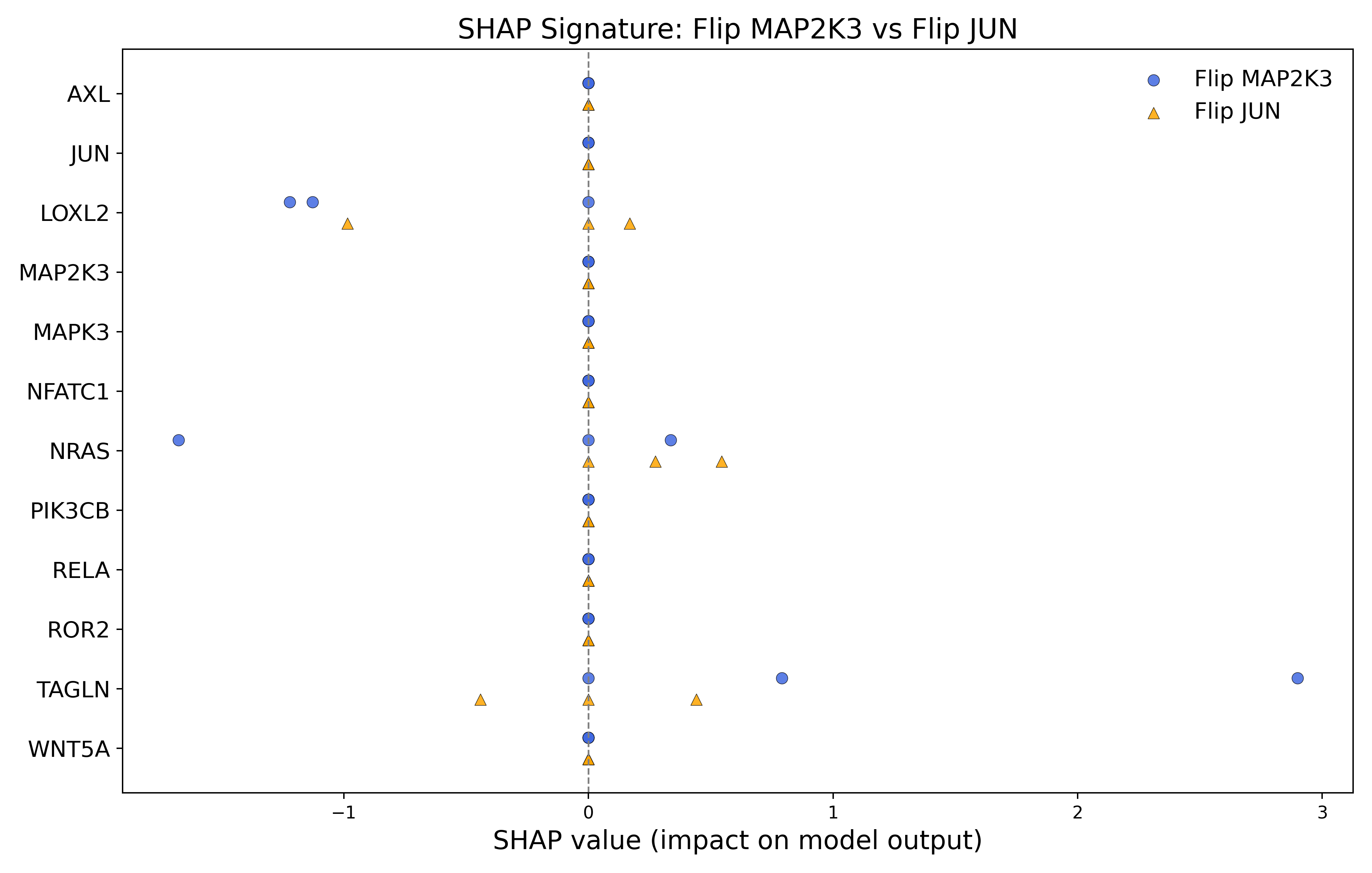}
	\caption{Comparative SHAP analysis for the agent's decision to flip MAP2K3 versus JUN from a resistant state. }
	\label{fig:shap_comparative_actions}
\end{figure}

\subsubsection{Context-Dependent Intervention Choice}
SHAP analysis reveals why the agent prioritizes MAP2K3 over JUN through context-dependent logic (Fig.~\ref{fig:shap_comparative_actions}). While high LOXL2 states favor flipping either MAP2K3 or JUN, NRAS and TAGLN states determine the choice: high NRAS directs MAP2K3 targeting, while high TAGLN favors JUN intervention.

This strategy aligns with melanoma biology. In NRAS-mutant contexts (15-20\% of melanomas \cite{ascierto2012role}), the agent targets MAP2K3 to intercept MAPK cascade activation \cite{domingues2018possible} at a strategic chokepoint. Conversely, when TAGLN drives epithelial-mesenchymal transition \cite{tsui2019transgelin}, the agent targets JUN \cite{hess2004ap} to disrupt resistance-associated transcriptional programs. This demonstrates pathway-specific intervention selection mirroring precision oncology strategies \cite{tsimberidou2020review}.

\subsubsection{Validation of the ``Hit-and-Run'' Mechanism}
\begin{figure}[t]
	\centering
	\includegraphics[width=\columnwidth]{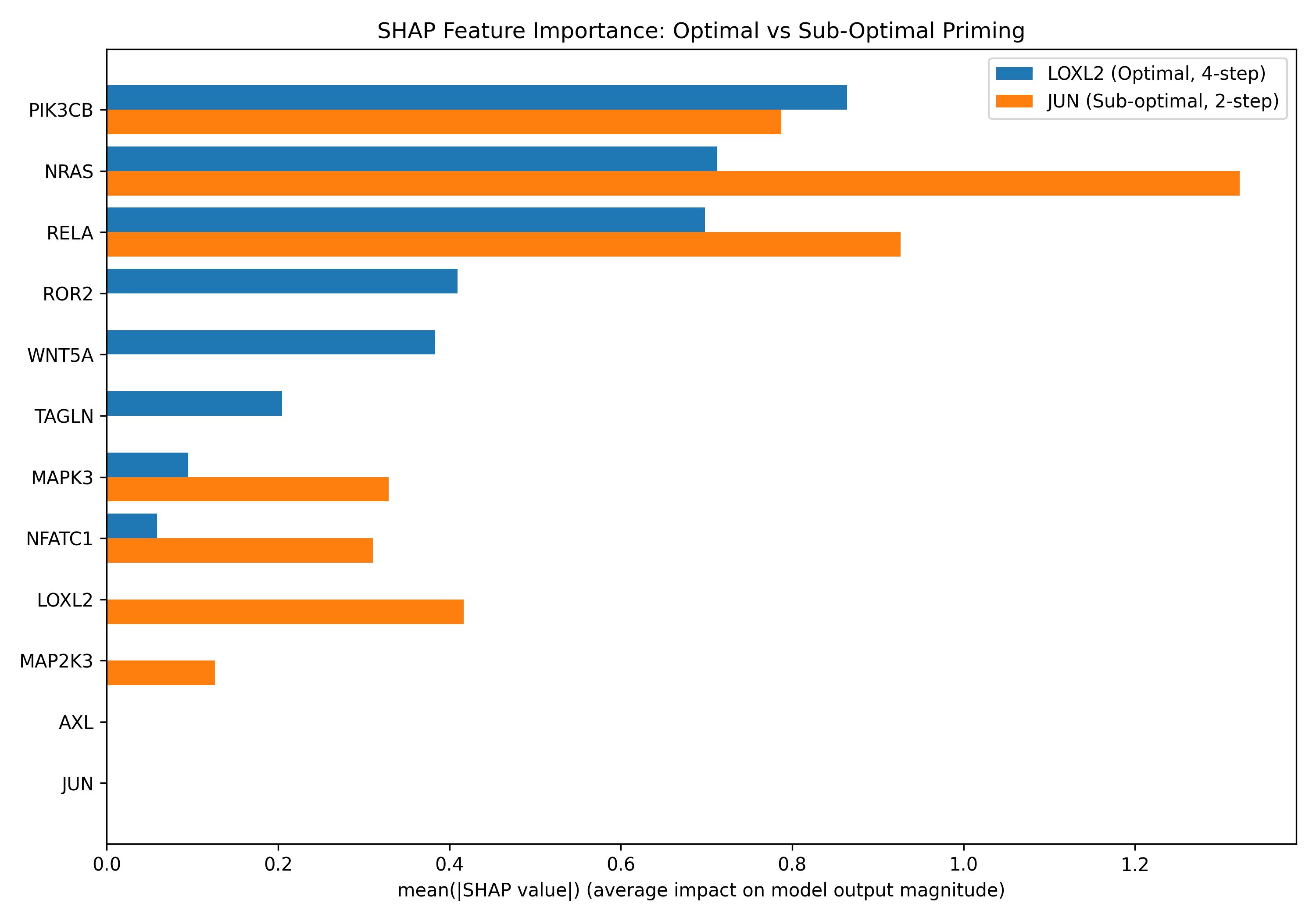}
	\caption{Residual SHAP importance after optimal (4-step LOXL2) versus sub-optimal (2-step JUN) priming. Effective LOXL2 priming requires less subsequent intervention.}
	\label{fig:shap_hit_run_validation}
\end{figure}

SHAP analysis explains why 4-step LOXL2 inhibition outperformed 2-step JUN inhibition by measuring residual ``vulnerability signatures'' (Fig.~\ref{fig:shap_hit_run_validation}). LOXL2 priming more effectively erases intervention triggers, leaving significantly lower SHAP values for key drivers like NRAS and RELA, indicating the network reached a more stable, less-resistant state.

LOXL2's superior efficacy reflects its pleiotropic role beyond matrix crosslinking—it acts as transcriptional regulator promoting epithelial-mesenchymal transition and coordinates multiple resistance pathways including hypoxia response and immune evasion\cite{barker2012rationale}. This aligns with its position as a master coordinator of resistance-associated programs\cite{tracy2016extracellular}.

\subsubsection{Dynamic Visualization of the Control Strategy}
\begin{figure}[t]
	\centering
	\includegraphics[width=\columnwidth]{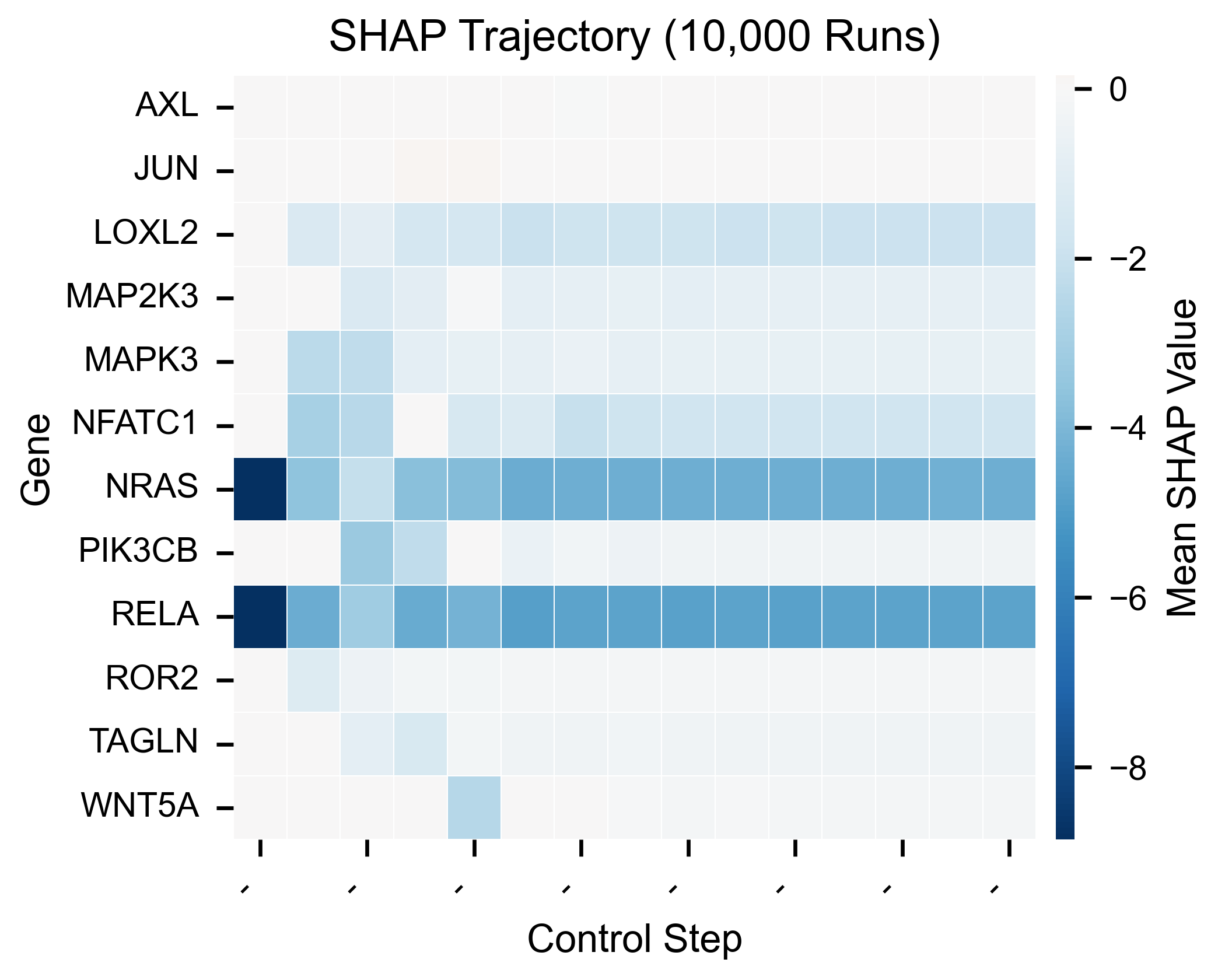}
	\caption{SHAP trajectory analysis showing persistent vulnerability signatures. ``Do Nothing'' was most frequent action, with NRAS/RELA negative signals that agent learns to ignore, validating the ``hit-and-run'' strategy.}
	\label{fig:robust_shap_trajectory}
\end{figure}

Analysis of 10,000 simulated episodes reveals persistent ``vulnerability signatures'' driving the agent's policy (Fig.~\ref{fig:robust_shap_trajectory}). The most frequent action was consistently ``Do Nothing,'' providing quantitative support for ``hit-and-run'' control where networks self-correct after initial priming.

The heatmap shows powerful initial negative SHAP contributions from NRAS and RELA (-0.4 to -0.6), indicating strong intervention impetus. This signal persists at lower intensity (-0.2 to -0.4) throughout episodes, suggesting these genes exert background pressure the agent learns to ignore. Other genes contribute weaker, transient signals before becoming neutral.

This temporal self-organization pattern aligns with successful clinical strategies like induction chemotherapy and immune checkpoint blockade, where initial perturbation triggers sustained therapeutic response. The identification of NRAS and RELA as persistent vulnerability drivers reflects established understanding of RAS/MAPK and NF-$\kappa$B as critical resistance mechanisms\cite{hugo2016genomic}.

\section{Conclusion}
This study introduced a novel computational framework integrating PBN modeling, reinforcement learning, and explainable AI to deconstruct and control innate immunotherapy resistance in melanoma. Our analysis revealed that resistance corresponds to a fundamental system-level transition from a plastic, multi-stable network to a rigid, canalized landscape driven by a JUN/LOXL2 regulatory axis that locks the system into a robustly resistant state.

Most significantly, our reinforcement learning approach discovered a non-obvious ``hit-and-run" therapeutic strategy: a precisely timed, 4-step temporary inhibition of LOXL2 achieving over 93\% success \textit{in silico}. Explainable AI analysis demonstrated that this transient intervention erases the molecular vulnerability signature stabilizing resistance, allowing the network's intrinsic dynamics to complete the therapeutic transition.

Our findings have important limitations. The binary gene representation oversimplifies continuous expression dynamics and misses intermediate states. Our 12-gene focus, while IPRES-motivated, excludes tissue-specific factors and broader regulatory networks. The PBN framework's discrete time steps and Boolean logic may not capture biological stochasticity and temporal dynamics. The model lacks spatial considerations, metabolic constraints, and immune interactions critical for \textit{in vivo} efficacy. Additionally, zero action costs in our RL formulation could yield different policies under alternative reward structures. This framework has only been tested on a single melanoma dataset, requiring validation across diverse cancer types and genetic backgrounds.

Clinical translation requires systematic validation: experimental testing in cell lines and organoids to confirm LOXL2 timing effects, preclinical studies in xenograft models combining temporal inhibition with immunotherapy, and development of biomarkers for patient selection. Clinical trials must establish optimal timing protocols, design adaptive frameworks for schedule optimization, and carefully monitor safety and resistance emergence. Phase I studies should focus on pharmacodynamics and immune activation patterns.

The discovery of time-dependent optimal strategies challenges traditional sustained inhibition paradigms. While requiring extensive validation, the integrated PBN-RL-XAI pipeline offers a generalizable computational approach for discovering novel control strategies in complex, treatment-resistant diseases, providing a powerful framework for future therapeutic development.

	\section*{Acknowledgment}
	
	The author would like to express sincere gratitude to Professor Zhang Louxin for his valuable advice on potential improvements to this work. Special thanks are also extended to Professor Ching Wai Ki for his thorough internal review and constructive feedback.
	
	% Use the IEEEtran bibliography style file
	\bibliographystyle{IEEEtran}
	
	\bibliography{IEEEabrv,reference}

	\appendix

	\begin{figure*}[htbp]
		\centering
		\includegraphics[width=\textwidth]{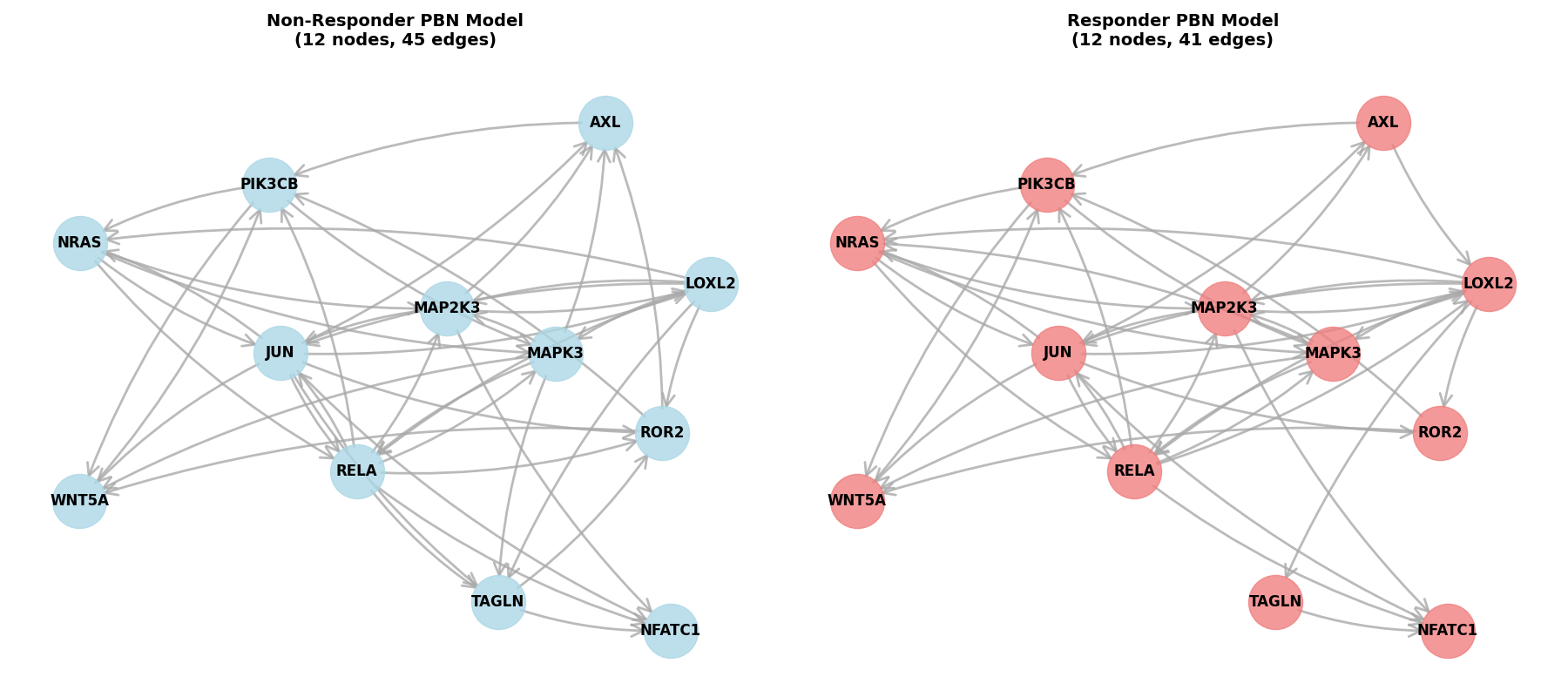}
		\caption{Comparative topology of responder and non-responder PBN models. Left: Non-responder network showing 45 regulatory edges with TAGLN functioning as a central hub. Right: Responder network with 41 edges and simplified regulatory architecture. Node colors distinguish the two phenotypes, with arrows indicating regulatory relationships.}
		\label{fig:network_topology}
	\end{figure*}
	
		\subsection{Code Availability}
	All source code developed for the PBN inference, reinforcement learning, and subsequent analyses presented in this paper are publicly available in our GitHub repository at \url{https://github.com/Liu-Zhonglin/pbn-melanoma-project}.

	\subsection{Network Topology Structure}
	
\begin{figure}[htbp]
		\centering
		\includegraphics[width=0.5\textwidth]{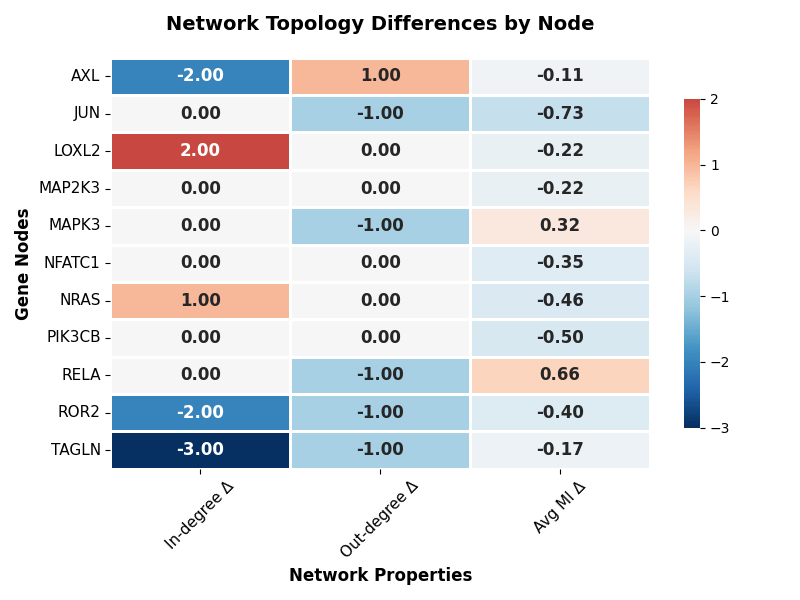}
		\caption{Network topology differences showing Responder - Non-responder values for nodes with significant topological changes.}
		\label{fig:topology_heatmap}
	\end{figure}

	Fig.~\ref{fig:topology_heatmap} quantifies topological differences between responder and non-responder PBN models, showing only nodes with substantial changes (absolute in-degree or out-degree differences $\geq$1, or average mutual information differences $\geq$0.2). The non-responder network exhibits higher connectivity (45 vs 41 edges, density 0.341 vs 0.311), suggesting increased regulatory complexity in resistant states. Blue regions indicate higher values in non-responder networks, while red indicates higher values in responder networks.
	
	\begin{table}[htbp]
		\caption{Network Topology Comparison for Nodes with Significant Changes}
		\begin{center}
			\begin{tabular}{|c|c|c|c|c|}
				\hline
				\textbf{Node}&\multicolumn{2}{|c|}{\textbf{In-degree}} & \multicolumn{2}{c|}{\textbf{Out-degree}} \\
				\cline{2-5} 
				& \textbf{\textit{Non-resp.}}& \textbf{\textit{Resp.}}& \textbf{\textit{Non-resp.}}& \textbf{\textit{Resp.}} \\
				\hline
				AXL & 4 & 2 & 1 & 2 \\
				\hline
				JUN & 4 & 4 & 8 & 7 \\
				\hline
				LOXL2 & 2 & 4 & 7 & 7 \\
				\hline
				TAGLN & 4 & 1 & 2 & 1 \\
				\hline
				ROR2 & 4 & 2 & 3 & 2 \\
				\hline
				\textbf{Total} & \textbf{45} & \textbf{41} & \textbf{45} & \textbf{41} \\
				\hline
			\end{tabular}
			\label{tab:topology_detailed}
		\end{center}
	\end{table}

	Key topology changes reveal distinct rewiring patterns: TAGLN shows the most dramatic changes with 3-fold reduced in-degree (-3) and decreased out-degree (-1) in responders, confirming its role as a resistance coordination hub that becomes disrupted in treatment-responsive networks. AXL shifts from signal integration (resistant) to active regulation (responsive) with increased out-degree (+1) and decreased in-degree (-2). LOXL2 gains regulatory inputs (+2) in responders, indicating enhanced ECM remodeling control, while ROR2 shows reduced connectivity (-2 in-degree, -1 out-degree), suggesting diminished Wnt pathway coordination in sensitive states.

\end{document}